\documentstyle[12pt,epsfig]{article}                                
\begin{document}

\titlepage                  

\begin{center}
\begin{large}
\begin{bf}

Dual degeneracy in supersymmetry

\end{bf}
\end{large}
\end{center}

 Biswanath Rath

\vspace{0.1cm}

Department of Physics,
 Maharaja Sriram Chandra Bhanj Deo University,
 Takatpur, Baripada -757003, Odisha, INDIA

(*biswanathrath10@gmail.com)

$\bf{Abstract:}$

We construct a double degenerate supersymmetry in one dimensional quantum mechanics. Here energy levels satisfy the conditions $E_{0,1}^{(-)}=0$ and 
$E_{n,n+1}^{(+)} = E_{n+2,n+3}^{(-)}$. The corresponding SUSY Hamiltonians($H^{(\pm)}$) are double degenerate in nature.The method has been tested on power law potentials.For harmonic oscillator model, we present analytical expression for energy $E_{n}^{(-)}$.
\vspace{0.1cm}

\begin{bf}
PACS no-03.65.Ge,11.30.Pb
\end{bf}

\vspace{0.1cm}

\begin{bf}
Key Words :
\end{bf}

double degenerate, supersymmetry,unbroken spectra,superpotential,double zero energy levels.

\begin{bf}
I.Introduction
\end{bf}

 In traditional supersymmetry, energy levels are non-degenerate in nature and satisfy the the relations[1-16] 
\begin{equation}
E_{n}^{(+)}=E_{n+1}^{(-)}
\end{equation}
and 
\begin{equation}
E_{0}^{(-)}= 0
\end{equation}
Here the corresponding SUSY Hamiltonians are $H^{(\pm)}$.

\begin{table}[htbp]
\begin{center}

\begin{bf}
Standard form of SUSY
\end{bf}

\vspace{1.0cm}

\begin{tabular}{  c c  } \hline \hline
 $H^{(-)}$  &$\hspace{1.0in}$ $H^{(+)}$   \\ \hline
----------  n=0&$\hspace{1.0in}$  \\
---------- n=1&$\hspace{1.0in}$-----------n=0  \\ 
---------- n=2&$\hspace{1.0in}$ ----------n=1  \\
---------- n=3&$\hspace{1.0in}$-----------n=2   \\ \hline \hline
\end{tabular}
\end{center}
\end{table}

\begin{bf}
II.Double generate quantum system
\end{bf}

In general,quantum 
Hamiltonians are degenerate in nature in higher dimensions.In this letter we 
propose a model to generate  degenerate Hailtonians in one dimension as follows.

Let $A$ is an annihilation operator 

\begin{equation}
A= \frac{d}{dx} + x
\end{equation}
so that ground state wave function is exactly known 

\begin{equation}
\Psi_{0}^{(-)}\sim e^{-x^{2}/2}
\end{equation}

Now we have to select the operator(s) $B$ as follows 

\begin{bf}
1. Quadratic model DDSUSY
\end{bf}

\begin{equation}
B = -\frac{d}{dx} + x -\frac{1}{x}
\end{equation}

The corresponding SUSY Hamiltonians are 
\begin{equation}
H_{1}^{(-)}= BA = p^{2}+x^{2}-2-\frac{1}{x}\frac{d}{d x}
\end{equation}
\begin{equation}
H_{1}^{(+)}= AB = p^{2}+x^{2}-\frac{1}{x}\frac{d}{d x} + \frac{1}{x^{2}}
\end{equation}
Out of the two $H_{1}^{(-)}$ is comparatively simple to handle.Using the Harmonic oscillator wave functions we get 

\begin{equation}
<\phi_{n}|\frac{1}{x}\frac{d}{dx}|\phi_{n}>= -1   ;n=0,2,4.....
\end{equation}

\begin{equation}
<\phi_{n}|\frac{1}{x}\frac{d}{dx}|\phi_{n}>= 1   ;n=1,3,5,.........
\end{equation}

Hence it is easy to see that 
\begin{equation}
<\phi_{n}|H_{n}|\phi_{n}>= 2n    ; n=even 
\end{equation}

\begin{equation}
<\phi_{n}|H_{n}|\phi_{n}>= 2n-2    ; n=odd 
\end{equation}

Here, $\phi_{n}$ is given by 
\begin{equation}
\phi_{n}=N_{n}H_{n}(x) e^{-x^2}/2)
\end{equation}
where $H_{n}$ stands for the Hermite Polynomial.

It is easy to see that the energy levels of $E_{n}^{(-)}=0,0;4,4;8,8;12,12;$

\begin{bf}
2.Quartic model DDSUSY
\end{bf}

Here, we just change $B$ inly as 

\begin{equation}
B = -\frac{d}{dx} + x^{3} -\frac{1}{x}
\end{equation}
The corresponding SUSY Hamiltonians are 
\begin{equation}
H_{2}^{(-)}= BA = p^{2}+x^{2}-2--x\frac{d}{d x}+x^{3}\frac{d}{d x}- \frac{1}{x}\frac{d}{d x}
\end{equation}
\begin{equation}
H_{2}^{(+)}= AB = p^{2}+x^{4}+3x^{2}-x\frac{d}{d x}+x^{3}\frac{d}{d x}- \frac{1}{x}\frac{d}{d x}+\frac{1}{x^{2}}
\end{equation}

In table-1, we display first few energy levels of above Hamiltonians as:

\begin{table}[htbp]
\begin{center}
 Table-1 :New double degenerate  SUSY

\vspace{1.0cm}
\begin{tabular}{ c   c     c  } \hline \hline
K& $H^{(-)}$&$\hspace{1.0in}$      $H^{(+)}$   \\ \hline
1&0---- n=0&$\hspace{1.0in}$ \\
 &0---- n=1&$\hspace{1.0in}$  \\
 &4---- n=2&$\hspace{1.0in}$4 n=0  \\
 &4-----n=3&$\hspace{1.0in}$4 n=1  \\
 &8---- n=4&$\hspace{1.0in}$8 n=2  \\
 &8-----n=5&$\hspace{1.0in}$8 n=3  \\ \hline
3 &0----n=0 &$\hspace{1.0in}$  \\
  &0-----n=1&$\hspace{1.0in}$   \\
  &6.635 8----  n=2&$\hspace{1.0in}$6.635 8---- n=0  \\
  &6.635 8----  n=3&$\hspace{1.0in}$6.635 8---- n=1  \\
  &16.497 8---- n=4&$\hspace{1.0in}$16.497 8----n=2  \\
  &16.497 8---- n=5&$\hspace{1.0in}$16.497 8----n=3  \\ \hline \hline
\end{tabular}
\end{center}
\end{table}

\begin{table}[htbp]
\begin{center}
\begin{tabular}{  c c  } \\
 quadratic operator $H_{1}^{(-)}$ & quartic operator $H_{2}^{(-)}$ \\
\includegraphics[height=2.0in,width=1.5in]{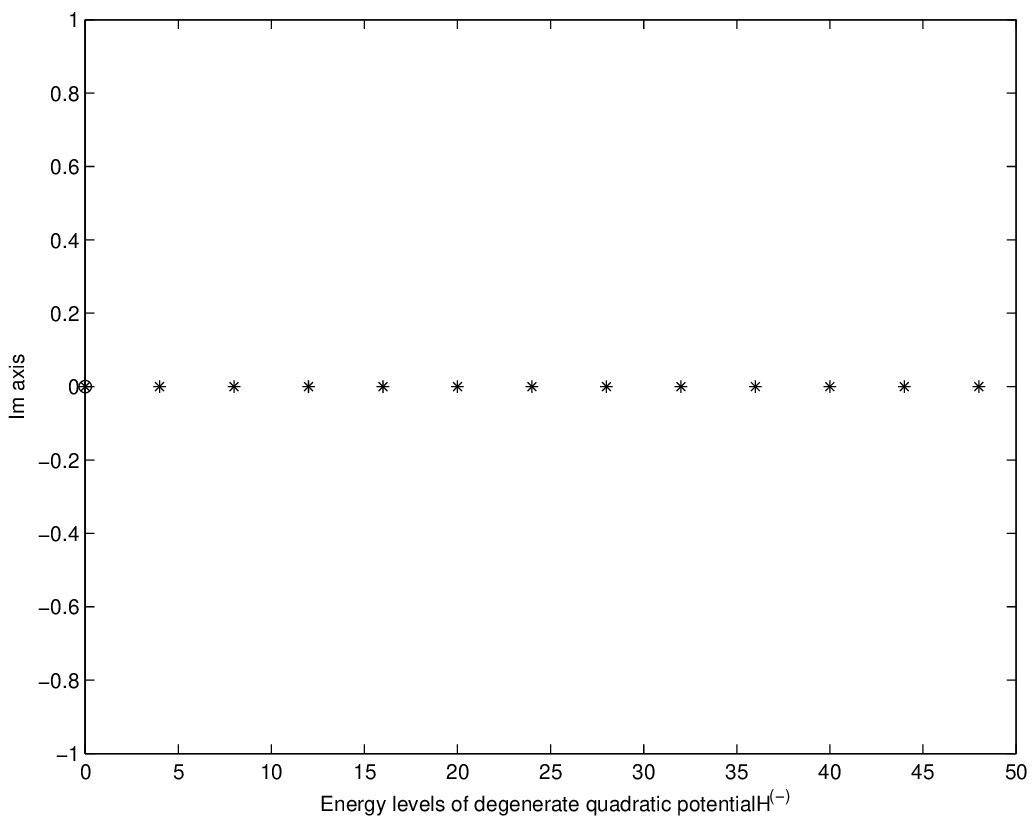} &
\includegraphics[height=2.0in,width=1.5in]{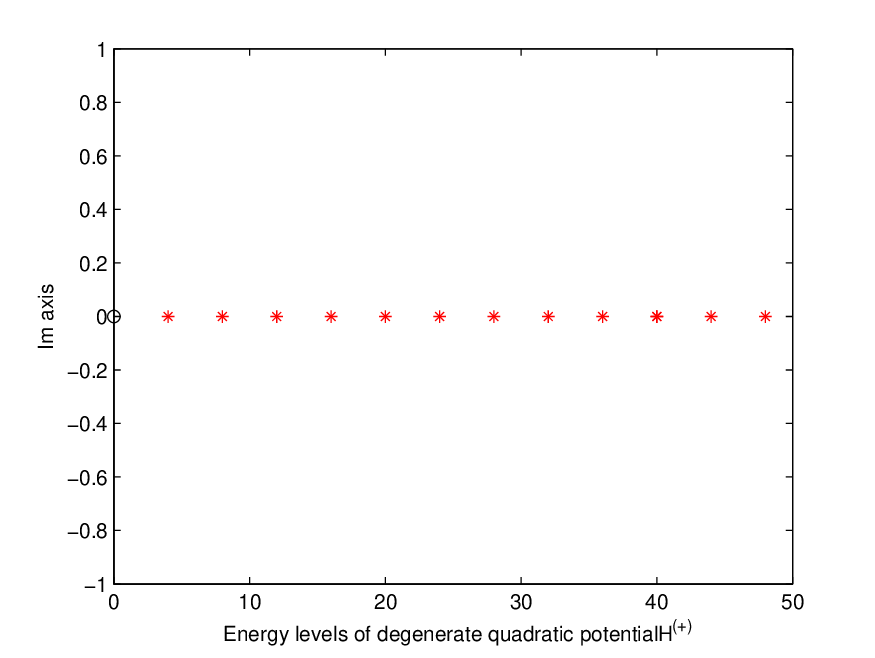}
\end{tabular}
\end{center}
\end{table}

In general,quantum operators can be generated from $B$ as  
\begin{equation}
B = -\frac{d}{dx} + x^{K} -\frac{1}{x}  \hspace{1.0cm} K=1,3,5,7
\end{equation}

\begin{bf}
III.Method of calculation 
\end{bf}

Here we use the matrix diagonalization method [14,15] to compute energy levels on solving the eigenvalue relation

\begin{equation}
H|\Psi>=E|\Psi>
\end{equation}
where 
\begin{equation}
|\Psi>=\sum A_{m} A_{m}|\phi_{m}>
\end{equation}
with 
\begin{equation}
H_{0}|\phi_{m}>=[p^{2}+x^{2}|\phi_{m}>=(2m+1)|\phi_{m}>
\end{equation}
In fact, explicit form of $\phi_{m}$ is given above.

\begin{bf}
IV.Conclusion
\end{bf}

In conclusion, we suggest a double degenerate unbroken SUSY see fig-2,3 for quadratic model potentials. For $H_{1}^{(-)}$ we present analytical result.Our numerical results also confirms this.For numerical results we use simple matrix diagonalisation method for the numerical evaluation of enerty levels.
Further the operators are complex in nature, so one need not be interested to see the wave function behaviour. On the other hand one can easily see the $|\Psi|^{2}$. For example, we plot the wave function mod square of the  first two states of $H^{(\pm)}$ of quadratic model potential. We believe the  degenerate SUSY can explore new physics in other branches of physics,preferably in particle physics[17].

\begin{table}[htbp]
\begin{center}
\begin{tabular}{  c c  } \\
 $|\Psi_{0,1}|^{2}$ : $H_{1}^{(-)}$ &  $H_{1}^{(+)}$ \\
\includegraphics[height=2.0in,width=1.5in]{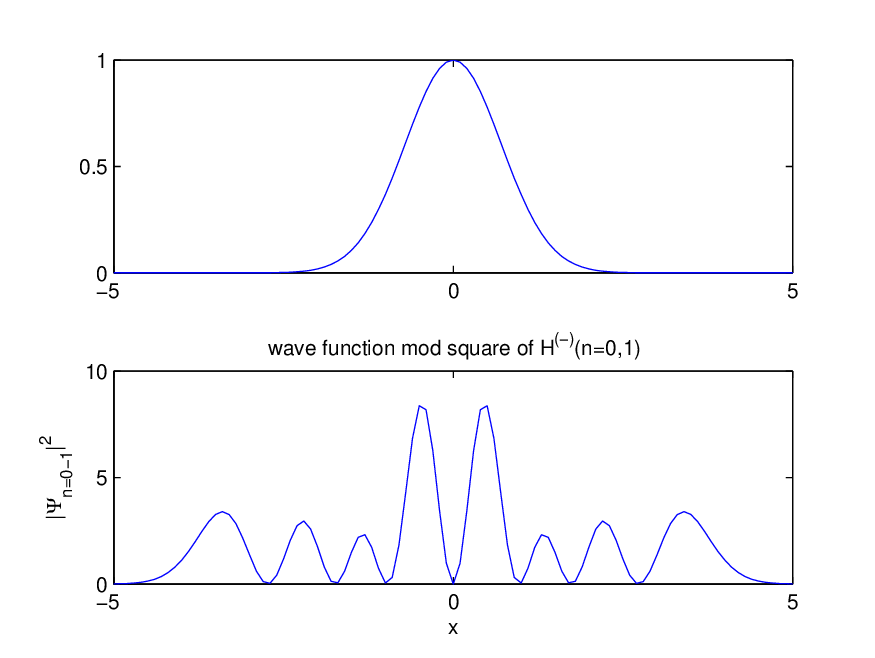} &
\includegraphics[height=2.0in,width=1.5in]{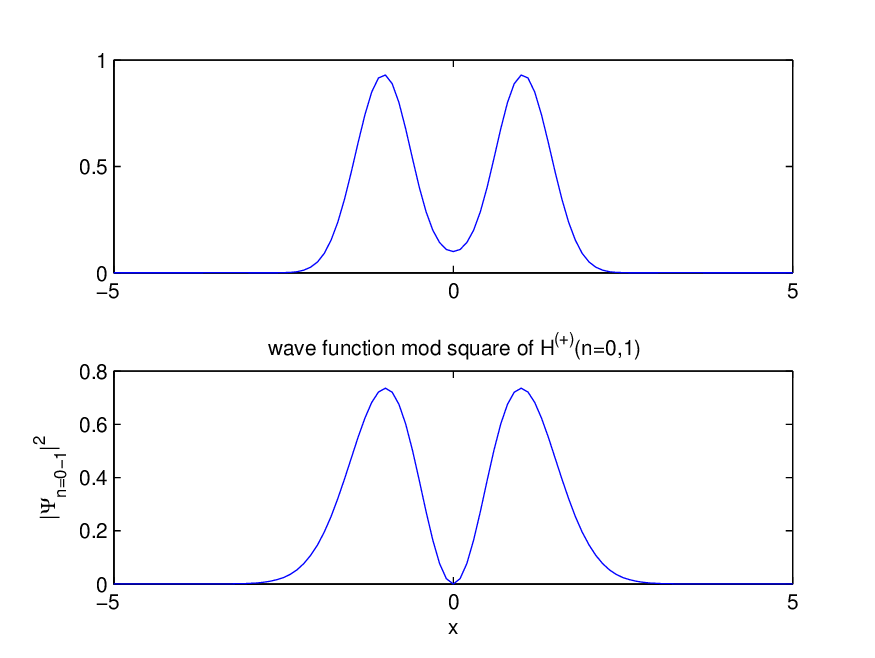}
\end{tabular}
\end{center}
\end{table}

 Lastly the present calculation has been done using the matrix diagonalisation method[18]. In condensed form degenerate SUSY can be depicted as 

\begin{table}[htbp]
\begin{center}
\begin{tabular}{  c c  } 
 $H^{(-)}$  &$\hspace{1.0in}$ $H^{(+)}$   \\ 
----------  n=0&$\hspace{1.0in}$ \\
----------  n=1&$\hspace{1.0in}$  \\ 
----------  n=2&$\clubsuit$ \hspace{1.0in} ----------n=0$\clubsuit$  \\
----------  n=3&$\clubsuit$ \hspace{1.0in}-----------n=1 $\clubsuit$ \\
\end{tabular}
\end{center}
\end{table}

\pagebreak

\end{document}